\documentclass[twocolumn, showpacs, preprintnumbers, amsmath, amssymb, nofootinbib]{revtex4}

\usepackage{graphicx}
\usepackage{dcolumn}
\usepackage{hyperref}
\usepackage[mathscr]{eucal}

\newcommand{\be}{\begin{equation}}
\newcommand{\en}{\end{equation}}
\newcommand{\bea}{\begin{eqnarray}}
\newcommand{\ena}{\end{eqnarray}}

\newcommand{\vc}[1]{\mbox{\boldmath$#1$}}
\newcommand{\bra}{\langle}
\newcommand{\ket}{\rangle}

\newcommand{\ssvc}[1]{\mbox{\scriptsize\boldmath$#1$}}
\newcommand{\tr}{\mbox{tr}\,}

\newcommand{\kb}{{\boldsymbol k}}
\newcommand{\pbb}{{\boldsymbol p}}
\newcommand{\D}{\mathscr{D}}
\newcommand{\hbo}{\hbox to 1 true cm {\hfill } }

\newcommand{\nn}{\nonumber}
\newcommand{\gt}{h}
\newcommand{\sutwo}{\mathop{\mathrm{SU}}(2)}

\begin{document}

\title{Coulomb gauge Gribov copies and the confining
potential}

\author{Thomas Heinzl}
\email{theinzl@plymouth.ac.uk}

\author{Kurt Langfeld}
\email{klangfeld@plymouth.ac.uk}

\author{Martin Lavelle}
\email{mlavelle@plymouth.ac.uk}

\author{David McMullan}
\email{dmcmullan@plymouth.ac.uk}

\affiliation{School of Mathematics and Statistics, University of
Plymouth\\
Drake Circus, Plymouth PL4 8AA, UK}

\date{\today}

\begin{abstract}
We study the approach, initiated by Marinari et al., to the
static inter-quark potential based on Polyakov lines of finite
temporal extent, evaluated in Coulomb gauge. We show that, at
small spatial separations, the potential can be understood as
being between two separately gauge invariant colour charges. At
larger separations Gribov copies obstruct the non-perturbative
identification of individually gauge invariant colour states.
We demonstrate, for the first time, how gauge invariance can be
maintained quite generally by averaging over Gribov copies.
This allows us to extend the analysis of the Polyakov lines and
the corresponding, gauge invariant quark-antiquark state to all
distance scales. Using large scale lattice simulations, we show
that this interpolating state possesses a good overlap with the
ground state in the quark-antiquark sector and yields the full
static inter-quark potential at all distances. A visual
representation of the Gribov copies on the lattice is also
presented.
\end{abstract}



\maketitle

\section{Introduction}
The inter-quark potential is a principal tool in lattice studies
of quark confinement \cite{Wilson:1974sk}. This potential can be
understood as composed of anti-screening and screening effects. In
perturbation theory the anti-screening structures are paradigm
effects of non-Abelian gauge theories and their dominance
underlies asymptotic freedom. Screening effects lower the energy
but, in perturbative calculations, are seen to be small compared
to anti-screening. Non-perturbatively, however, screening due to
light quarks starts to dominate at sufficiently large distances
implying string breaking. Hence, as was stressed for example  in
\cite{Gross:1995bp}, confinement with a linearly rising potential
at large inter-quark distances is a feature only of Yang-Mills
theory with no light quarks.

Although the potential itself must be gauge invariant, it is
initially unclear how to construct it since the heavy fermions are
not gauge invariant. On the lattice one measures the vacuum
expectation value of the gauge invariant Wilson loop and its
large-time behaviour \cite{Wilson:1974sk},
\be \label{WLOOP}
  W_{rT} \sim \exp \left\{ - T V(r) \right\} \; ,
\en
where $r$ and $T$ denote the spatial and temporal extent of the
loop. Wilson's formulation is based on connecting the fermions by
a path ordered exponential to produce a single gauge invariant
object. However, this is not unique and in this paper we will
study other ways of constructing gauge invariant probes of the
potential.

To clarify our notation we denote states constructed from the
heavy fermionic fields of the theory by $| q_{\ssvc{x}}
\bar{q}_{\ssvc{y}} \ket $. These are not physical states as they
are not gauge invariant. Gauge invariant quark-antiquark states
will be denoted by $| Q_{\ssvc{x}} \bar{Q}_{\ssvc{y}} \ket $.

\begin{figure}
\begin{center}
\includegraphics[width=6cm]{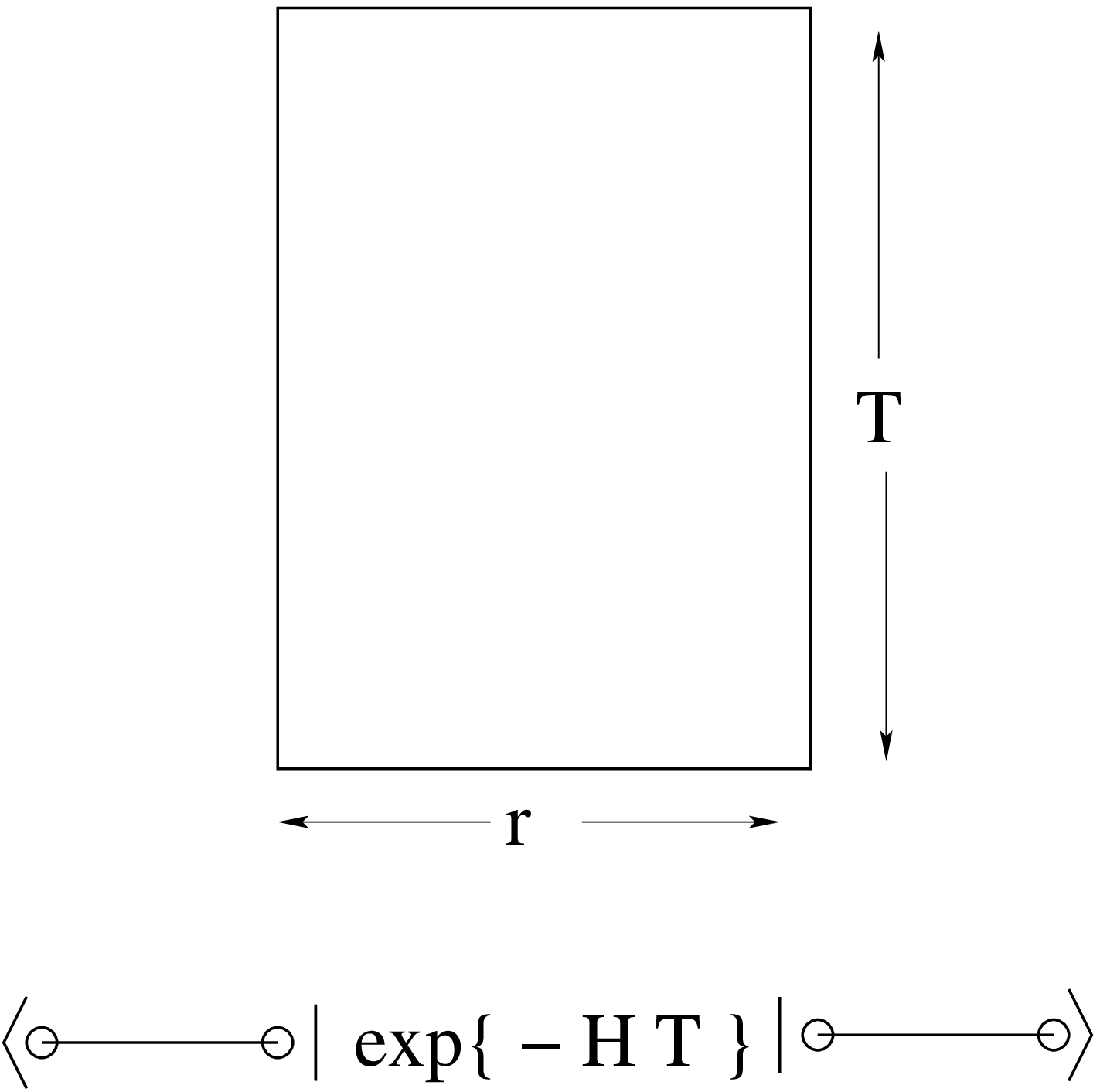}
\caption{\label{fig:wp1} The static inter-quark potential from unsmeared
Wilson loops.}
\end{center}
\end{figure}

As discussed above, the most familiar way to ensure gauge
invariance is to link the fermions by a Wilson line. Then the
Wilson loop (\ref{WLOOP}) may be found from the transition
amplitude
\be \label{TRANSAMP}
  W_{rT} \sim \bra \bar{Q}_{\ssvc{y}} Q_{\ssvc{x}}| e^{-HT} |
  Q_{\ssvc{x}} \bar{Q}_{\ssvc{y}} \ket \; , \quad r \equiv |\vc{x} -
  \vc{y}| \; ,
\en
which describes the creation of a static quark-antiquark pair
at $t=0$ and its subsequent annihilation at time $t=T$, see
FIG.~\ref{fig:wp1}.

Inserting a complete set of energy eigenstates, (\ref{TRANSAMP})
becomes at large times
\be \label{MOLOOP}
  W_{rT} \sim  \left| \bra \bar{Q}_{\ssvc{y}} Q_{\ssvc{x}} |
  0_{\ssvc{x}\ssvc{y}} \ket
  \right|^2 \exp \left\{ -  T V(r) \right\} \; ,
\en
where $ |   0_{\ssvc{x}\ssvc{y}} \ket $ signifies the lowest
energy state with a heavy quark and antiquark fixed at $\vc{x}$
and $\vc{y}$. For practical computational purposes it is of course
important that the matrix element in front is sufficiently large
or, put differently, that the state $ |Q_{\ssvc{x}}
\bar{Q}_{\ssvc{y}} \ket$ has a substantial overlap with the true
quark-antiquark ground state. Gauge invariance alone does not
ensure this and, in particular,   Wilson loops do not have such a
strong overlap. This is remedied on the lattice by the technique
of `smearing'~\cite{Albanese:1987ds,Teper:1987wt,Bali:1992ab}.

Although Wilson loops are gauge invariant  and hence
potentially physical they \emph{cannot} be interpreted as
amplitudes connecting \emph{separately} physical, \emph{single}
quark and antiquark states. This is symbolically displayed in
FIG.~\ref{fig:wp1} by the lines connecting the external sources
(small circles). However, in the perturbative sector physically
distinguishable constituent structures are
expected~\cite{Lavelle:1995ty}. To incorporate this constituent
picture into the description of the potential we would require
a gauge invariant quark state $|Q_{\ssvc{x}}\ket$ built out of
the heavy, but gauge dependent, fermionic state
$|q_{\ssvc{x}}\ket$.

Such an approach was, in fact, initiated by
Dirac~\cite{Dirac:1955uv} many years ago in QED and during the
last decade or so this idea has been developed into a fully
fledged alternative approach to charges in gauge
theories~\cite{Lavelle:1995ty,Horan:1998im,Usannals,Ilderton:2007qy}.
The process of constructing a gauge invariant quark from the
matter states $|q_{\ssvc{x}}\ket$ is called `dressing' and details
of the construction will be summarised in the next section. We
will see that the dressing for a static charge takes its simplest
form in Coulomb gauge. In that gauge, the fermionic state
$|q_{\ssvc{x}}\ket$ itself captures the dominant gluonic
contribution to the perturbative quark constituent of our
quark-antiquark configuration. As such, the state $|q_{\ssvc{x}}
\bar{q}_{\ssvc{y}}\ket$ evaluated in Coulomb gauge has a large
overlap with the true quark-antiquark ground state,
$|0_{\ssvc{x}\ssvc{y}} \ket$, at short distances.

\begin{figure}
\begin{center}
\includegraphics[width=6cm]{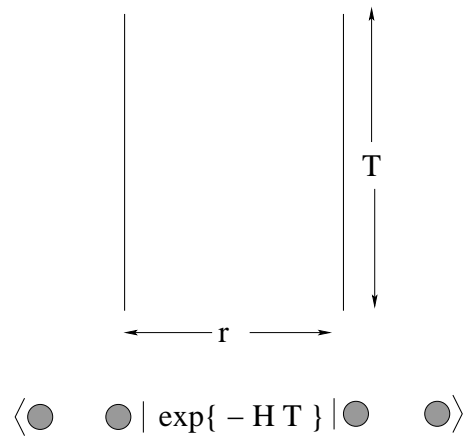}
\caption{ \label{fig:wp1d} The static inter-quark potential from dressed
quarks.}
\end{center}
\end{figure}

This observation allows us to develop an alternative to the
unsmeared (or \lq thin\rq) Wilson loop description of the
quark-antiquark system. We can now make the gauge invariant
identification that for heavy static quarks,
\be\label{QQSTATED}
  |Q_{\ssvc{x}}
  \bar{Q}_{\ssvc{y}}\ket\left|\right._{\partial_iA_i=0} =
  |q_{\ssvc{x}} \bar{q}_{\ssvc{y}} \ket\,.
\en
This is gauge invariant as long as each orbit of the gauge
group has one unique representative satisfying the Coulomb
gauge condition. Thus the expectation value of the Hamiltonian
in the state~(\ref{QQSTATED}) should give an alternative
approach to the inter-quark potential, see FIG.~\ref{fig:wp1d}.

An immediate question, which we will address in this paper, is to
what extent does the good perturbative overlap found in Coulomb
gauge extend to the non-perturbative regime? If it does, then, as
well as producing an alternative to smearing, we can start to
address issues related to the details and range of validity of the
constituent quark picture.

In this paper we will test this approach to the inter-quark
potential. That is, we will use Coulomb gauge fermions to
measure the potential. On a lattice this means that we first
take a Coulomb dressed  $Q\bar{Q}$ state on a time slice
separated by a distance $r$ and let it evolve during a time
interval $T$, see FIG.~\ref{fig:wp1d}. This yields a correlator
of two separate finite length Polyakov lines with link
variables in Coulomb gauge. The correlation of finite length
Polyakov lines was first studied on the lattice by Marinari et
al.~in~\cite{Marinari:1992kh}. They considered SU(3) gauge
theory and obtained an upper limit for the string tension which
overestimated the true value by roughly 50\%. It was recently
confirmed (for an SU(2) gauge theory) that finite length
Polyakov line correlators indeed yield the correct value for
the full string tension~\cite{Greensite:2003xf}. One of the
goals of this paper is to apply this approach not just to the
string tension but to the full inter-quark potential. Using
both perturbation theory and lattice gauge simulations, we will
find that the potential between two Coulomb dressed quarks and
that obtained from Wilson loops are in good agreement.

An intriguing aspect of the present investigation is the obvious physical
fact that the constituent picture has a limited range of validity.
Thus any interpretation of the inter-quark potential as that
arising between two physical quarks must break down. It has been
argued in references~\cite{Lavelle:1995ty,Ilderton:2007qy} that
this is consistent with the observation that the gauge invariance
of the dressed charges breaks down non-perturbatively due to
Gribov copies~\cite{Gribov:1977wm,Singer:1978dk}. In this regime
the potential defined by the finite length Polyakov lines can only
be interpreted as arising from the  state which corresponds to one
gauge invariant, but overall colourless, object. However, it is
not \emph{a priori} obvious that the residual gauge invariance of
the states~(\ref{QQSTATED}) due to copies will let us extract a
potential, let alone the confining one. Therefore, another aim of
the work reported here is to study how Coulomb gauge Gribov copies
affect the inter-quark potential. One might wonder whether summing
over such copies could mean that the potential vanishes (as the
construction is not gauge invariant with respect to
transformations between Gribov copies). In fact we shall see that
we can extract the potential in this way.

The structure of this paper is as follows. First we shall discuss
what is known about the perturbative inter-quark potential. This
will include a discussion of the relevant ingredients of the
dressing approach to static charges and a summary of their use in
identifying the screening and anti-screening structures found in
the potential. This will allow us to clarify the significance of
Coulomb gauge for the description of static quarks. Then, in
section~3, the non-perturbative regime will be investigated on the
lattice. Results from the use of unsmeared Wilson loops will be
compared with those that arise from finite length Polyakov lines
in Coulomb gauge. In section~4 we will then investigate in more
detail the Gribov copies that arise in this approach and present
arguments and evidence for why the inter-quark potential is
insensitive to such copies. We will end with some comments and
conclusions.

\section{The perturbative potential}

\subsection{Dressing approach}

In order to be physical charges must be invariant under gauge
transformations which map $\vc{A} \to {^\Omega} \! \vc{A}$ and
$|q_{\ssvc{x}}\ket \to \Omega_{\ssvc{x}} |q_{\ssvc{x}}\ket$. The
generic form for a charged single-fermion state is thus given
by\footnote{Note that in order to conform to lattice conventions,
the gauge transformations have been renamed $h \to h^\dagger$
etc.\ as compared
to~\cite{Lavelle:1995ty,Horan:1998im,Usannals,Ilderton:2007qy}.}
\be \label{QINV}
  |Q_{\ssvc{x}} \ket \equiv h_{\ssvc{x}} [\vc{A}] \, |q_{\ssvc{x}}
  \ket \; ,
\en
where the field dependent dressing, $h_{\ssvc{x}} [\vc{A}]$, must
transform as,
\begin{equation}\label{ginv}
  h_{\ssvc{x}} \big[{^\Omega} \!\vc{A} \big] = h_{\ssvc{x}}
  [\vc{A}] \, \Omega_{\ssvc{x}}^\dagger \, ,
\end{equation}
for $|Q_{\ssvc{x}} \ket$ to be a gauge singlet. Note that the
dressing $h[\vc{A}]$ can be viewed as a field dependent gauge
transformation. The invariance of (\ref{QINV}) from (\ref{ginv})
is a minimal requirement and there are many ways to construct such
invariant states. The simplest manner to write down a gauge
invariant state involving a fermion-antifermion pair at different
points $\vc{x}$ and $\vc{y}$ is to connect them by a straight
Wilson line. This is, however, not physically attractive and, as
the need for smearing on the lattice shows, is also not the best
thing to do in practice.

At a more fundamental level, the problem with such a description
may already be seen in QED~\cite{Haagensen:1997pi} where the
potential between two static fermions, separated by a distance
$r=|\vc{x}-\vc{y}|$ and connected (dressed) by a \textit{string},
is easily calculated to be proportional to $e^2 r \,
\delta^{2}(\vc{0})$. This is a linearly rising potential (with a
divergent coefficient) which would imply confinement in QED.
However, upon quantisation the associated state is clearly
infinitely excited and thus energetically unphysical. On a
discrete lattice the coefficient would be finite, but this would
still correspond to a highly excited state which will decay. The
reason why the energy is so high is that the electric field is
only non-zero on the path of the Wilson line, e.g.\
\be
  \vc{E} (\vc{x}) = \delta^{2} (\vc{x}_\perp) \theta (x_3) \,
  \vc{n} \; ,
\en
where $\vc{n}$ is the unit vector in the $x_3$-direction and
$\vc{x}_\perp = (x_1 , x_2)$.
The large energy of such thin strings is exactly the reason why
smearing Wilson lines is numerically efficient in lattice
calculations of the inter-quark potential: it improves the overlap
with the ground state
\cite{Albanese:1987ds,Teper:1987wt,Bali:1992ab}.

Instead of this we look for a physically motivated description of
the dressing of two static charges\footnote{The generalisation to
moving charges is discussed in detail in~\cite{Usannals}}. We
demand that a dressed state is gauge invariant and static as an
asymptotic state~\cite{Horan:1998im,Usannals}. This produces a
dressing which factors into two parts:
\begin{equation}\label{sa}
  h = e^{i\kappa} e^{i \chi} \, .
\end{equation}
For a static charge, $\chi$ is the field dependent transformation
into Coulomb gauge. This part of the dressing  we also call the
`minimal' dressing since the state $e^{i \chi(\ssvc{x})}
|q_{\ssvc{x}} \ket$ is gauge invariant on its own. Hence, its
$Q\bar{Q}$ analogue,
\be
  | Q_{\ssvc{x}}\bar{Q}_{\ssvc{y}} \ket \equiv e^{i \chi(\ssvc{x})} e^{-i
  \chi(\ssvc{y})} |q_{\ssvc{x}} \bar{q}_{\ssvc{y}} \ket \; ,
\en
may be expected to have a large overlap with the true $Q\bar{Q}$
ground state $| 0_{\ssvc{x} \ssvc{y}} \ket$ (involving the full
dressing). The additional part of the dressing, $\kappa$, is the
time integral of the $\chi$ rotated $A_0$ component of the
potential. So in Coulomb gauge it is just the (long) Polyakov line
of $A_0$ from the current time to infinity.

It turns out to be a nontrivial task to find a closed expression
for $\chi$, the transformation into Coulomb gauge in the
non-Abelian case. It is, however, simple to
obtain a perturbative expansion for $\chi$ to high orders.
Absorbing the coupling, $g$, into the vector potential, the
minimal dressing at leading order (LO) is
\be
\chi=\frac{\partial_iA_i}{\nabla^2}+{\cal O}(g^2)\,.
\en
In the Abelian limit this is the result proposed by
Dirac~\cite{Dirac:1955uv} to describe a static electron. At
next-to-leading order (NLO) and beyond we have
\begin{equation}\label{gthree}
  \chi = \chi_1 + \chi_2 + \chi_3 + {\cal O}(g^4)\,,
\end{equation}
where explicit expressions for the $\chi_n$s can be found
in~\cite{Lavelle:1995ty,Ilderton:2007qy}. Given this solution for
$\chi$, it is then reasonably straightforward to construct the
additional part of the dressing to the same order in the coupling
by using $\chi$ to gauge rotate $A_0$. This perturbative
construction can be extended to an arbitrary order, however, the
above suffices to calculate the NLO potential.

A fermionic field dressed by (\ref{sa}) has two independently gauge
invariant structures: $\kappa $ and the minimally dressed fermion.
Given these separately gauge invariant terms, it is instructive to
investigate their relative contributions to the potential.
The potential between two such dressed quarks is
found by sandwiching the Yang-Mills Hamiltonian between states of
a dressed quark and antiquark spatially separated by a distance
$r$. As in QED, the LO dressing generates the Coulombic electric
field typical of short distances~\cite{Lavelle:1995ty} and the
additional part of the dressing makes no contribution here.  So to
LO we find
\be \label{LOPOT}
  V_{_{\mathrm{LO}}}(r) = \bra \bar{Q}_{\ssvc{y}} Q_{\ssvc{x}}| H_{\mathrm{YM}} | Q_{\ssvc{x}}
  \bar{Q}_{\ssvc{y}} \ket - E_0  = \frac{g^2 C_F}{4\pi r} \; .
\en
Here, $E_0$ collectively denotes contributions independent of the
distance $r$. As usual, $C_F$ denotes the quadratic Casimir in the
fundamental representation. The result (\ref{LOPOT}) corresponds
to one-gluon exchange at tree level as will be briefly discussed
in the following subsection.

At NLO we focus on the minimally dressed quark where we only
retain the Coulomb dressing. The higher order terms in the minimal
dressing~(\ref{gthree}) modify the essentially Abelian result
(\ref{LOPOT}) and generate the paradigm anti-screening effects
which underly asymptotic freedom. In momentum
space~\cite{Lavelle:1998dv}, working in an arbitrary covariant
gauge\footnote{Working in an arbitrary covariant gauge in this way
allows us to check the gauge invariance of the construction. It
also allows us to avoid the need for Dirac bracket commutators
that arise in, for example, Coulomb gauge. } in $d$~spatial
dimensions, the chromo-electric commutators produce at NLO the
correction
\begin{equation}\label{Vg4}
  -\frac{3g^4C_FC_A   k_i k_j}{\kb^4}\int\frac{d^dp}{(2\pi)^d} \frac1{ (\kb - \pbb)^2}
  iD_{ij}^{TT}(\pbb) \,,
\end{equation}
where $iD_{ij}^{TT}(\pbb) $ is the gauge invariant two-point
function of the transverse components $\vc{A}^T$ of the spatial
vector potential and $C_A$ is the quadratic Casimir in the adjoint
representation. The potential from the minimal dressing at this
order in perturbation theory thus becomes
\begin{equation}\label{Vg4min}
  V^{^{\mathrm{min}}}_{_{\mathrm{NLO}}}(\vc{k}) = \frac{g^2 C_F}{\vc{k}^2} \left\{ 1 + \frac{g^2
  C_A}{48 \pi^2} \, 12  \ln \left(\frac{\mu^2}{\vc{k^2}}\right) \right\} \; ,
\end{equation}
where $\mu$ is a renormalisation scale. The factor of 12 in front
of the logarithm demonstrates that the minimal dressing
description yields the dominant \textit{anti-screening} part of
the quark potential which is thus generated by the glue needed to
make quarks gauge invariant. The expected factor of $11$ in the
true ground state (familiar from the beta function) requires an
additional $-1$ term which is produced by screening due to gauge
invariant glue
--- recall that the additional dressing, $\kappa$, is gauge
invariant, see~\cite{Bagan:2001wj}. The sign of this smaller effect
shows that it lowers the overall energy.

In three dimensions~\cite{Bagan:2000nc} the minimal dressing
generalises this anti-screening/screening divide in the
inter-quark potential despite the fact that this
super-renormalisable lower dimensional theory has a vanishing
beta-function. It is noteworthy that the relative weighting of the
divide at NLO is almost identical in both three and four
dimensions.  The renormalisation of these physical descriptions of
charges has been investigated at NNLO and a cancellation of
non-local divergences was shown~\cite{Bagan:2005qg}. This was a
highly non-trivial check of the construction.

We note also that the arbitrary path dependence of $Q\bar{Q}$
states dressed by a Wilson line may be factorised
order by order in perturbation theory. This yields a product
dressing consisting of a gauge invariant, but path dependent,
gluonic term and a separately gauge invariant structure of two
fermions dressed with the above minimal (or Coulomb)
dressing~\cite{Lavelle:1999ki}.

\subsection{Wilson loop approach}

To make the transition from the dressing description to the
lattice formulation, it is helpful to briefly recapitulate the
perturbative route to the potential in terms of Wilson loops,
Polyakov lines and Feynman graphs as initiated
in~\cite{Susskind:1976pi} and reviewed in \cite{Kogut:1982ds}. A
path integral definition of the static inter-quark potential
$V(r)$ may be given as follows. We consider the partition function
\be \label{eq:ZRHO}
  Z[\rho] = \int \D A \, \exp \Bigl\{ - \left[ S_{\mathrm{YM}} + (\rho, A^0)
  \right] \, \Bigr\}
\en
in the presence of a static external $q\bar{q}$ source (at
$\vc{x}$ and $\vc{y}$, respectively),
\be
  \rho_{\ssvc{x}\ssvc{y}}^a (\vc{z}) = -g T^a [\delta(\vc{z} -
  \vc{x}) - \delta (\vc{z} - \vc{y})] \; .
\en
The measure in $(\ref{eq:ZRHO})$ implicitly contains gauge fixing
terms as does the action $S_{\mathrm{YM}}$. The partition function
$Z$ is the exponential of the Schwinger functional,
\be
  Z[\rho] \equiv \exp \left\{- i W[\rho] \right\} \; ,
\en
which yields the static potential in the large-time limit, $T \to
\infty$, plus constant self-energy contributions $\Sigma$,
\be \label{eq:WRHO}
  W[\rho] \to T \, [ V(\vc{r}) + \Sigma ] \; .
\en
If we define the untraced Polyakov line
\be
  P_T (\vc{x}) \equiv \mathrm{T} \exp \left\{-i g
  \int_{0}^{T} dt \, A_0 (t, \vc{x})  \right\} \; ,
\en
we may rewrite the amplitude (\ref{eq:ZRHO}) as an expectation
value, or more properly a two-point function, given by
\be \label{eq:ZRHOPP}
  Z[\rho] = \bra \tr P_T(\vc{x}) P_T^\dagger (\vc{y}) \ket
  = \int \D A \, \tr P_T(\vc{x}) P_T^\dagger (\vc{y})
  \, e^{ - \,  S_{\mathrm{YM}}} \; .
\en
In contrast to the dressing approach to the potential,  gauge
invariance is not manifest here since the nonlocal operator $\tr
P_T(\vc{x}) P_T^\dagger (\vc{y})$ transforms nontrivially as the
Polyakov lines do not trace a closed loop. However, for some gauge
choices (e.g.\ Landau or Coulomb gauge) one expects the spatial
part of the gauge potential to decay sufficiently fast in temporal
direction  that one can close the integration contours to form a
rectangular Wilson loop~\cite{fischler:1977}. Choosing spatial and
temporal extent $r$ and $T$, respectively, the resulting Wilson loop
(expectation value) is
\be
  W_{rT} \equiv \left\langle \tr \mathrm{P} \exp \left\{- ig \oint_{rT} dx_\mu A^\mu
  \right\} \right\rangle \sim \exp \left\{ - T V(r) \right\} \; ,
\en
with the final expression being approached in the large time
limit, $T \to \infty$. From this familiar gauge invariant
construction,  perturbative calculations in \textit{covariant}
gauges of the potential have been performed to two
loops~\cite{peter:1997b,schroder:1998}.

To connect with the minimal dressing approach, it is useful to
recall the calculation of the perturbative inter-quark potential
in Coulomb gauge
\cite{khriplovich:1969,duncan:1976,frenkel:1976,appelquist:1977}.
Taking into account that the Coulomb gluon propagator has a
transverse and an instantaneous part, we have, at NLO,
\be \label{eq:POTEXP}
\includegraphics{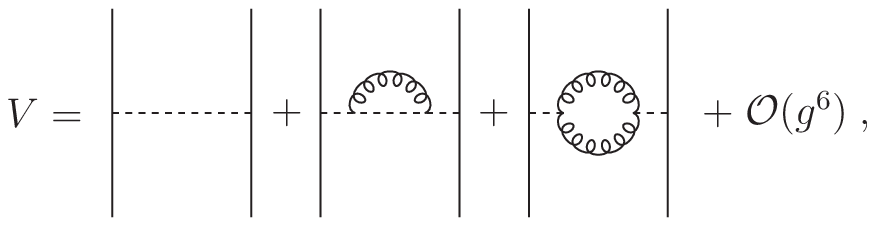}
\en
where the dotted lines represent `unphysical' instantaneous
Coulomb gluons, $A_0$, with propagator
\be
  D_{00} (\vc{k}) = \frac{i}{\vc{k}^2} \; ,
\en
while the `curly' lines correspond to `physical' transverse gluons
$A_i^T$ with propagator
\be
  D_{ij}^{TT} (k^0, \vc{k}) = \frac{i}{k_0^2 - \vc{k}^2 + i \epsilon} \left(
  \delta_{ij} - \frac{k_i k_j}{\vc{k}^2} \right) \; .
\en
The full vertical lines represent the heavy-quark propagators or,
equivalently, the untraced Polyakov lines.

Looking at (\ref{eq:POTEXP}) we note that only gluonic two-point
insertions contribute. The second diagram corresponds to the
minimal contribution to the potential~(\ref{Vg4}) while the third
diagram is the smaller screening contribution and may be traced
back to the separately gauge invariant dressing, $\kappa$, of
(\ref{sa}).

Higher $n$-point functions will only contribute at two-loop order
and above in the coupling. We mention in passing that this order
has never been worked out in Coulomb gauge as one is faced with a
number of difficulties. First, there are severe infrared
singularities \cite{appelquist:1978}. Second, integrations over
energies ($p^0$) tend to diverge badly. Third, there are
nontrivial contributions from the Faddeev-Popov measure and the
Christ-Lee terms \cite{christ:1980b,nachbagauer:1995} induced by
the curvature of the configuration space. Items two and three seem
to be related \cite{Cheng:1986hv,doust:1987a,doust:1987b}.

Working out the diagrams of (\ref{eq:POTEXP}) the potential in
momentum space becomes
\be
  V_{_{\mathrm{NLO}}} (\vc{k}) = \frac{g^2 C_F}{\vc{k}^2} \left\{ 1 + \frac{g^2
  C_A}{48 \pi^2} (12 - 1) \ln \frac{\mu^2}{\vc{k^2}} \right\}
   \; ,
\en
where the quantity in brackets determines the running coupling.
Asymptotic freedom arises because the \textit{anti-screening} or
minimal-dressing contribution yields a factor $+12$, cf.~(\ref{Vg4min}),
as compared to the $-1$ from the \textit{screening}
vacuum polarisation term.

One can understand the different signs at NLO by noting that
`physical' particles (like the transverse gluons) in a loop always
entail screening by unitarity \cite{halzen:1984,dokshitzer:2004}.
At small distances this effect is due to virtual particles in the
loop. At large distances, however, screening effects may be caused
by real particles. This is the case for string breaking in QCD
where light dynamical quark pairs pop out of the vacuum and
combine with the heavy probe quarks to form heavy-light mesons.
This in turn leads to a saturation of the potential at long
distances. The same happens in pure gluodynamics if the heavy
sources are in higher group representations such that they can be
screened by dynamical gluons.

Beyond NLO the separation into screening and anti-screening
structures is not so straightforward and the only systematic
approach is through the dressing decomposition~(\ref{sa}). So, for
example, at NNLO physical particles in loops can now contribute to
anti-screening~\cite{Bagan:2005qg}. The gauge transformation
properties of the dressing ensures that, at all orders in
perturbation theory, the separation of forces into screening and
anti-screening ones is gauge invariant and hence physical. How
this is achieved in the non-perturbative sector is an open
question.

\section{The non-perturbative regime}

The use of Coulomb gauge in the non-perturbative regime has
received much attention recently: there have been
investigations  using lattice techniques
\cite{Cucchieri:2000gu,Cucchieri:2000kw,Langfeld:2004qs,Greensite:2003xf,
Greensite:2004ke,Nakamura:2005ux,Nakagawa:2006fk},
variational arguments
\cite{Zwanziger:2002sh,Szczepaniak:2001rg,Szczepaniak:2003ve,Feuchter:2004mk},
Schwinger-Dyson equations \cite{Fischer:2005qe} and explicit
analytic solutions~\cite{Ilderton:2007qy}.
In particular the so-called Coulomb potential has been
studied in great detail. This potential arises from the short Polyakov line correlator~(\ref{eq:ZRHOPP}) in the limit
$T \to 0$~\cite{Greensite:2003xf,Greensite:2004ke,Nakamura:2005ux,Nakagawa:2006fk}
and must not be confused with the full static potential, which we address
in the present study.
In this section we shall recall the key ingredients of the lattice
approach and then see how to extract the inter-quark potential
from finite length Polyakov lines.

\subsection{Numerical setup \label{sec:cg} }

The dynamical degrees of freedom are the unitary matrices
$U_\mu (x)  \in  \sutwo$ which are associated with the links
of a $N^4$ cubic lattice with lattice spacing $a$.
The partition function is given by
\be \label{eq:i3}
  Z  =  \int \mathscr{D} U \; p[U] \; , \quad \mathrm{with} \quad
  p[U]  =  \exp \left\{  \beta \; S[U] \right\} \, ,
\en
where the Wilson coupling $\beta $ is the only free parameter of
the simulation. It multiplies the gauge invariant Wilson action,
\be
  S[U] = \frac{1}{2} \, \sum _{\mu<\nu, x} \tr \; U_\mu (x) \; U_\nu (x+\mu) \;
  U^\dagger _\mu (x+\nu) \; U^\dagger _\nu (x) \, ,
\en
which is used throughout this paper. The integration measure is
given in terms of single-site Haar measures $dU_\mu (x)$:
\be
  \mathscr{D} U  =  \prod _{x, \mu } dU_\mu (x) \, .
\en
The latter are invariant under left and right multiplication by
$\sutwo$ group elements and normalised to unity, $ \int dU_\mu (x)
= 1$. This ensures the gauge invariance of $\mathscr{D} U$ and,
therefore, of the partition function. Under a gauge transformation
$\Omega(x)$, the links change according to
\be
  U_\mu (x) \; \longrightarrow \, ^{\Omega } U _\mu (x)  =
  \Omega (x) \; U_\mu (x) \; \Omega ^\dagger (x+\mu ) \, .
\en
In order to impose Coulomb gauge in a lattice simulation, one
firstly determines a link dependent gauge transformation,
$\gt(x)$, from the `action principle',
\be
  S_\mathrm{fix}  =  \sum _{t, \ssvc{x}, i} \tr \;
  ^\gt U _i (t, \vc{x}) \; \stackrel{\gt}{\longrightarrow } \; \hbox{max} \, .
  \label{eq:gf1}
\en
which, on each gauge orbit, looks for the representative $^\gt U
_i$ that is closest to the unit matrix. Note that the sum over
times, $t$, guarantees that this is done term by term on each
time-slice.

Lattice configurations in Coulomb gauge are then made up from the
gauge transformed links $^\gt U _\mu (t, \vc{x})$. The search for
a maximum of the gauge fixing functional $S_\mathrm{fix}$ is
performed with a standard iteration and over-relaxation procedure.
For reasonable lattice sizes, present day algorithms will
typically find a local (rather than the global) maximum of the
gauge fixing action (\ref{eq:gf1}) on each orbit.

Starting with the same `seed' configuration $U_\mu (x)$, we might
end up in several, almost degenerate, maxima of (\ref{eq:gf1})
implying that the lattice definition of the Coulomb gauge
configuration is ambiguous. This is the lattice version of the
continuum Gribov ambiguity discussed in the introduction. Two
configurations $^{\gt_1} U_\mu (x) $ and $^{\gt_2} U_\mu (x) $
with both $\gt_1(x)$ and $\gt_2(x)$ satisfying (\ref{eq:gf1}) are
hence called Gribov copies of each other. Since we are primarily
interested in the effects induced by Gribov copies, we do not
employ sophisticated tools such as simulated annealing or Fourier
acceleration which are invoked (with little success it must be
said) to find the global maximum of (\ref{eq:gf1}).

\subsection{Confining potential from unsmeared Wilson loops }

The standard way to calculate the potential of a static
quark-antiquark pair is to investigate the transition amplitude
(\ref{TRANSAMP}), which we now write as
\be \label{eq:wp1}
  W (r,T) \equiv \bra \bar{Q}_{\ssvc{y}} Q_{\ssvc{x}}| e^{-HT} |
  Q_{\ssvc{x}} \bar{Q}_{\ssvc{y}} \ket \, ,
\en
where, as before, $H$ denotes the Yang-Mills Hamiltonian, $T$ a
Euclidean time interval and $r = |\vc{x} - \vc{y}|$ the separation
between quark and antiquark. Recall that the latter are connected
 by a parallel transporter along a straight line. We will evaluate
$W(r,T)$ via lattice Monte Carlo simulations employing
rectangular Wilson loops (see FIG.~\ref{fig:wp1}). As discussed
in the introduction, there are sophisticated lattice techniques
\cite{Albanese:1987ds,Teper:1987wt,Bali:1992ab} available for
enhancing the overlap, $\bra \bar{Q}_{\ssvc{y}}
Q_{\ssvc{x}}|0_{\ssvc{x}\ssvc{y}}\ket$, with the
quark-antiquark ground state by using (spatially) smeared
Wilson loops. However, since we are not only interested in the
static potential \emph{per se}, but also in the overlap of
different interpolating states with the true ground state, we
refrain from adopting these techniques and content ourselves
with a thorough comparison of unsmeared Wilson loops, where
there is only an overall gauge invariant quark-antiquark state,
with individually dressed Coulombic states as pictorially
summarised  in FIG.s \ref{fig:wp1} and \ref{fig:wp1d}.

Let us first discuss the evaluation of (\ref{eq:wp1}). Rewriting
the amplitude as
\be
  W (r,T)  \equiv  \exp \{ - T\; v(r,T)   \} \, , \label{eq:wp2}
\en
the inter-quark potential $V(r)$ emerges as the limit
\be
  V(r) = \lim _{T \to \infty } v(r,T) \; . \label{eq:wp3}
\en
\begin{figure}[h]
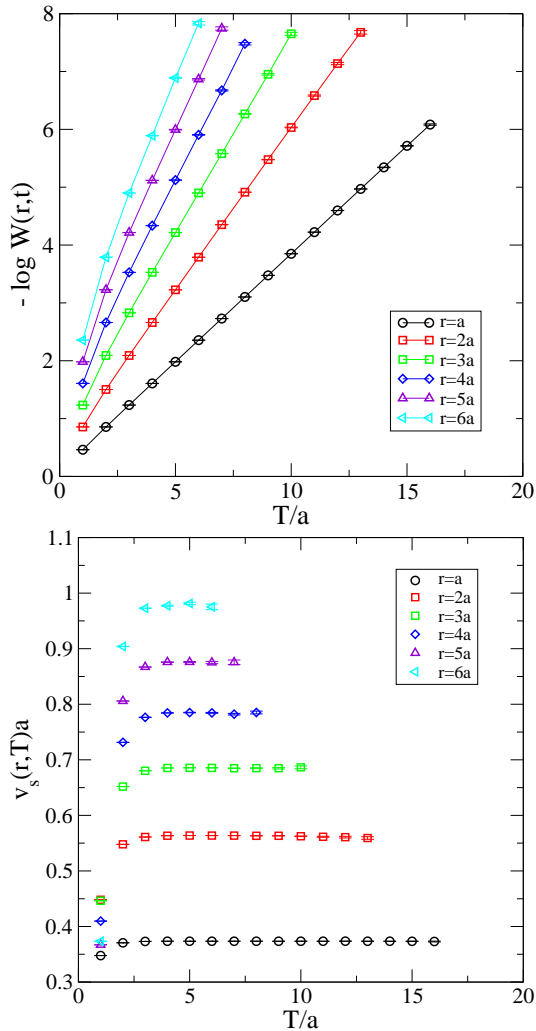


\includegraphics[width=7cm]{wilson_td}

\includegraphics[width=7cm]{wilson_t}

\caption{ \label{fig:wp2} The dependence of unsmeared Wilson
loops on $T$; $16^4$ lattice, $\beta=2.4$. }
\end{figure}

The large $T $ limit ensures that excited states which
contribute to the matrix element (\ref{eq:wp1}) are suppressed.
In this limit $v(r,T)$ becomes independent of $T$.
FIG.~\ref{fig:wp2} (upper panel) shows $- \ln W(r,T) = T
v(r,T)$ as a function of $T$ for several values of $r$.
Deviations from the linear behaviour are clearly visible at
small values for $T$. The curves for larger values of $r$ seem
to be more affected by excited states.

In practice, we fit the logarithm of (\ref{eq:wp2}),
\be
  -  \ln  W(r,T)  \equiv  b(r)  +  V(r)  T \label{eq:wp4}
\en
to a linear function in $T$ (we call this a $T$-fit). Only data
points with $T \ge T_{\mathrm{th}}$, for some threshold time
$T_{\mathrm{th}}$, are taken into account. Using small values for
the threshold time, say $T_{\mathrm{th}} = 2a$, the $\chi
_T^2/\mathrm{dof}$ becomes unacceptably large. This indicates
large deviations from linear behaviour and implies that the signal
is contaminated by contributions from excited states. We have
therefore made a more conservative choice for the threshold,
\be
  T_{\mathrm{th}}  =  4 \, a \; , \hbo \hbo (16^4 \;
  \hbox{lattice}, \; \beta = 2.4) \label{eq:wp5}
\en
and included only $T$-fits satisfying
\be
  \chi_T^2/\mathrm{dof} < 2 \; . \label{eq:wp6}
\en
The lower panel of FIG.~\ref{fig:wp2} shows
\be
  v_s (r,T) \equiv [ - \, \ln \; W(r,T) \; - \; b(r) ] / T = v(r,T) - b(r)/T
\en
as a function of $T$. It is reassuring to note that, for
$T>4a$, the contributions from excited states are within the
statistical error.
\begin{figure}[h]
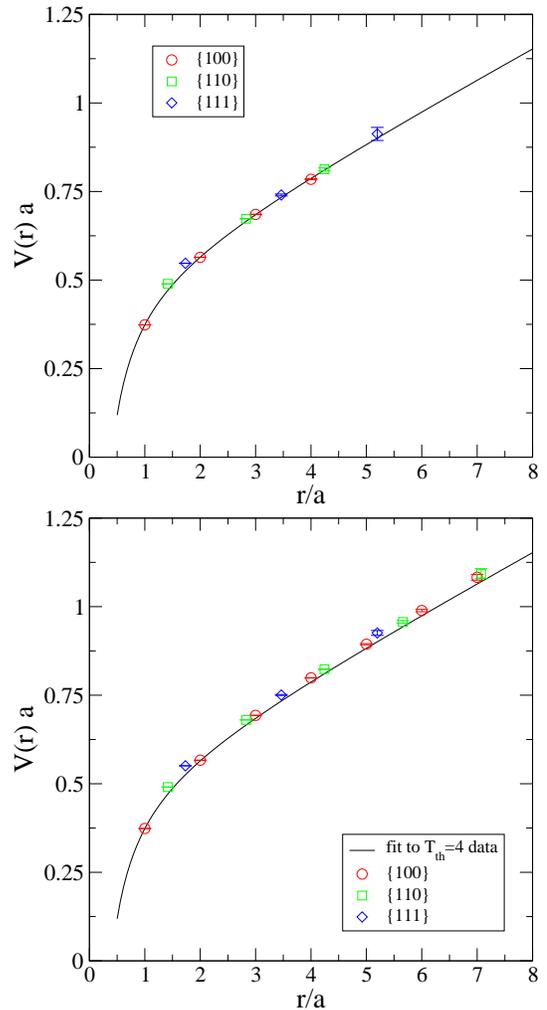


\includegraphics[width=7cm]{poten_wil4}

\includegraphics[width=7cm]{poten_wil3}

\caption{ \label{fig:wp3} The static potential extracted from
unsmeared Wilson loops as a function of the quark-antiquark
distance $r$ using $T_{\mathrm{th}}=4a$ (upper panel) and
$T_{\mathrm{th}}=3a$ (lower panel) as threshold for the
$T$-fit; $16^4$ lattice, $\beta=2.4$. }
\end{figure}
The $T$-fit described above also provides us with the desired
static inter-quark potential $V(r)$. Generically, the
quark-antiquark pair is aligned along the main crystallographic
direction, e.g.\ $\{100\}$, of the lattice. In order to check
for artifacts arising due to rotational symmetry breaking, the
spatial sides of the Wilson loop were placed along the diagonal
axes $\{110\}$ and $\{111\}$ as well. Our final result for the
unsmeared loop and $T_{\mathrm{th}} = 4a$ is shown in the left
panel of FIG.~\ref{fig:wp3}. To increase the number of
statistically significant data points for $V(r)$ we also
relaxed the condition (\ref{eq:wp6}) by using the threshold
$T_{\mathrm{th}}=3a$, see FIG.~\ref{fig:wp3}, lower panel. Also
included in this graph is the solid line from the fit to the
$T_{\mathrm{th}}=4a$ data. We find that $T_{\mathrm{th}}=3a$
data approach this line reasonably well. Note that the
contribution of the excited states in the $T_{\mathrm{th}}=3a$
data yields a slight overestimate of the string tension.

\vskip 0.3cm Finally, the potential $V(r)$ is fitted to the
universal function
\be
  V(r)  =  V_0  -  \frac{\alpha }{r}  +  \sigma \, r
  \, , \label{eq:wp7}
\en
with offset $V_0$, Coulombic coefficient $\alpha$ and string
tension $\sigma$ to be determined (we call this a $V$-fit). Our
findings are summarised in Table~\ref{tab:wp1}.
\begin{table}
\caption{ \label{tab:wp1} Fit of the static potential to the
function (\protect\ref{eq:wp7}) for a $16^4$ lattice and for
$\beta = 2.4$. }
\smallskip
\begin{ruledtabular}
\renewcommand{\arraystretch}{1.4}
\begin{tabular}{c|cccc}
& $V_0$ & $\alpha$  & $\sigma a^2$  & $\chi_V^2 / \mbox{dof}$  \\ \hline
$T_\mathrm{th}=3a$     & $0.487(1)$ & $0.204(1)$ & $0.0908(4)$ & $13.0$ \\
$T_\mathrm{th}=4a$     & $0.501(3)$ & $0.212(2)$ & $0.0847(8)$ & $4.7$\\
\end{tabular}
\end{ruledtabular}
\end{table}
Not unexpectedly, we find that caution is required when the
static potential is extracted from unsmeared Wilson loops:
There is a significant contribution from excited states the
suppression of which requires a sufficiently large Euclidean
time extent $T$. For the present settings employing a $16^4$
lattice and $\beta = 2.4$, we find a $7\%$ drop in the string
tension from $\sigma a^2 \approx 0.0908$ to $\sigma a^2 \approx
0.0847$ if we impose the more rigorous bound (\ref{eq:wp6}) for
the $T$-fit. Note also that calculations which use smeared
Wilson loops for an overlap enhancement find smaller values for
the string tension, see e.g.\ \cite{Fingberg:1992ju}. This
indicates that, even for the case of the more conservative
bound (\ref{eq:wp6}), there are probably still some
contributions from excited states.

\subsection{Confining potential from Coulomb dressed quarks }

In what follows we will employ lattice gauge configurations
which have been transformed to the Coulomb gauge along the
lines described in subsection~\ref{sec:cg}. Our working
hypothesis here is that we can extract the physical inter-quark
potential despite the presence of Gribov copies. The authors
of~\cite{Marinari:1992kh,Greensite:2003xf} found that the large
distance part of the potential, i.e., the string tension, is
not affected by Gribov copies. However, no physical explanation
of this observation was given. Although the copies obstruct the
\textit{interpretation} of the resulting potential as that
arising between separately invariant single quark sources
\cite{Lavelle:1995ty} we shall demonstrate that they do not
affect the overlap with the ground state.

The overlap of a Coulomb dressed, single heavy quark state at
$t=0$ with the analogous state at $t=T$ is given by a finite
length Polyakov line (evaluated in Coulomb gauge),
\be
  ^\gt P_T (\vc{x})  \equiv  \prod _{t=0}^{T} \, ^{\gt[U]}  U_0
  (\vc{x},t) \, , \label{eq:cp1}
\en
where $^{\gt} U_\mu (x)$ denotes the link after gauge fixing.
According to (\ref{eq:ZRHOPP}) the potential of two static Coulomb
dressed quarks can then be extracted from the correlator of two
Polyakov lines of finite temporal extent,
\be
  C(r,T)  \equiv  \Bigl\langle \; \tr \; {^\gt} \! P_T(\vc{x}) \;
  {^\gt} \! P^\dagger_T(\vc{y}) \; \Bigr\rangle \; . \label{eq:cp2}
\en
This correlator was studied in~\cite{Greensite:2003xf} with an
emphasis on the Coulomb (minimal anti-screening) potential obtained from
the limit $T \rightarrow 0 $. In this paper, we will concentrate
on the full static potential (large $T$ behaviour) and the impact of
Gribov copies.
For sufficiently large values of $T$ the excited states are
suppressed and we expect $C(r,T)$ to be dominated by the true
$Q\bar{Q}$ ground state $| 0_{\ssvc{x}\ssvc{y}} \ket $. Hence,
defining quantities analogous to (\ref{eq:wp2}) and
(\ref{eq:wp3}),
\be
  C(r,T) \equiv \exp \{ -T u(r,T) \} \; , \quad U(r) \; = \;
  \lim _{T \to \infty} \; u(r,T) \, ,
\label{eq:cp3}
\en
we expect $U(r)$ to coincide with the static potential $V(r)$
extracted from unsmeared Wilson loops. The purpose of the
present subsection is to scrutinise this expectation by a
detailed numerical investigation.

As we will have to deal with gauge fixing ambiguities let us first
address the issue of residual gauge invariance. As is appropriate
for a Hamiltonian formulation,  the Coulomb gauge
prescription~(\ref{eq:gf1}) is implemented on time slices, $t =
\mathrm{const}$. Hence, we may still perform spatially
homogeneous, purely time dependent residual gauge transformations,
$\Omega(t)$, on the gauge fixed links,
\be \label{RESGT}
  {^\gt} U_\mu (t, \vc{x}) \to {^{\Omega(t)\gt}} U_\mu (t, \vc{x}) = \Omega(t)
  \, {^\gt} U_\mu (t, \vc{x}) \;
  \Omega^\dagger (t) \, ,
\en
which, in continuum language, obviously leave the (vanishing)
divergence of $\vc{A}(t , \vc{x})$ invariant.

Let us now convince ourselves that the correlator $C(r,T)$ of
(\ref{eq:cp2}) is invariant under (\ref{RESGT}),
\begin{widetext}
\be \label{eq:cp4}
  \tr ^{\Omega(t)} \! P_T(\vc{x}) \,  ^{\Omega(t)} \! P_T^\dagger (\vc{y}) =
  \tr \Omega(0) \, P_T(\vc{x}) \, \Omega^\dagger(T) \, \Omega(T)
  \, P_T^\dagger (\vc{y}) \, \Omega^\dagger(0) =
  \tr P_T(\vc{x}) \, P_T^\dagger (\vc{y}) \, ,
\en
\end{widetext}
where the links can, in fact, be taken in any gauge. Thus,
$C(r,T)$, and with it the potential $U(r)$ from (\ref{eq:cp3}),
are not affected by the trivial residual gauge freedom
(\ref{RESGT}).

For a quark trial state which is not completely invariant because
of Gribov copies, one might naively expect that the average over
gauge copies causes the correlator $C(r,T)$ to vanish. Our numerical
simulations, however, show that
this does not happen: we obtain a clear signal for $C(r,T)$, which
is orders of magnitude larger than the underlying noise. We
therefore apply the linear $T$-fit to the function $-\ln C(r,T)$
subject to the rigorous bounds
\be
  T_\mathrm{th} \; = \; 4a \; , \hbo \chi _T^2/\mathrm{dof} < 2 \; .
  \label{eq:cp5}
\en
\begin{figure}[!]
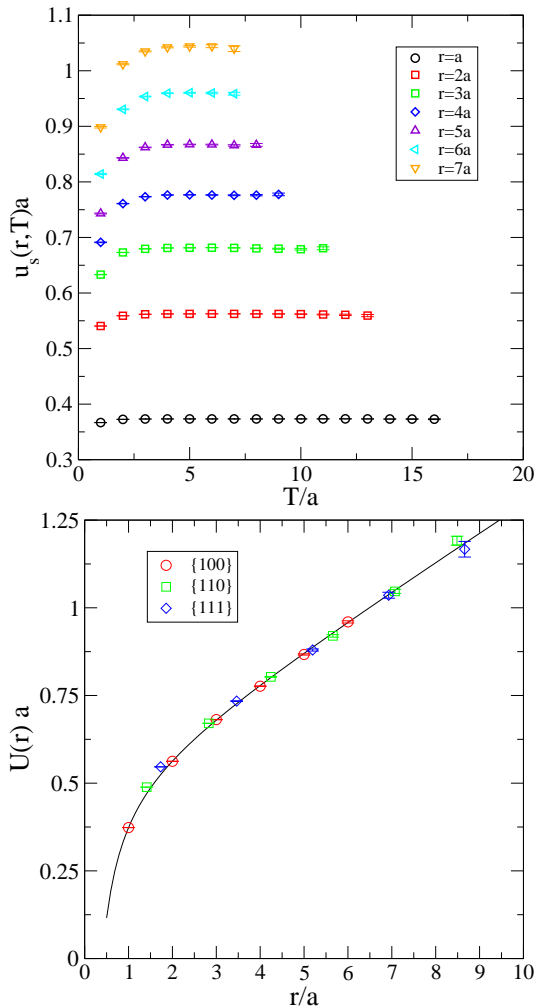


\includegraphics[width=7cm]{gribov_t}

\includegraphics[width=7cm]{poten_gri4}

\caption{ \label{fig:cp1} $u(r,T)$ extracted from the
correlation of finite length Polyakov lines calculated from
Coulomb gauge fixed lattice configurations (upper panel). The
static potential calculated from finite length Polyakov lines
(lower panel). Both on a $16^4$ lattice with $\beta=2.4$. }
\end{figure}

The fit results allow us to extract $u(r,T)$ to high accuracy
as shown in FIG.~\ref{fig:cp1}, upper panel. The outcome is
clearly encouraging: not only is the  Gribov noise absent, it
also appears that the overlap with the ground state is largely
enhanced. In FIG.~\ref{fig:cp1} even the $r=7a$ data are
statistically significant. This is also true for the plot of
the static potential $U(r)$ defined in (\ref{eq:cp3}). Finally,
we have again fitted the parametrisation (\ref{eq:wp7}) of the
potential to the present data (FIG.~\ref{fig:cp1}, lower
panel). The results of this $V$-fit are listed in
Table~\ref{tab:cp1}.
\begin{table}[h]
\caption{ \label{tab:cp1} Fit of the static potential $U(r)$ to
the function (\ref{eq:wp7}) for a $16^4$ lattice and for $\beta
= 2.4$. }
\begin{ruledtabular}
\renewcommand{\arraystretch}{1.5}
\begin{tabular}{c|cccc}
       & $V_0$ & $\alpha $ & $\sigma a^2 $ & $\chi_V^2 / \mathrm{dof} $ \\ \hline
$T_\mathrm{th}=3a $    & $0.505(1)$ & $0.214(1)$ & $0.0827(2)$ & $11.0$ \\
$T_\mathrm{th}=4a $    & $0.510(2)$ & $0.217(1)$ & $0.0807(4)$ & $6.5$
\end{tabular}
\end{ruledtabular}
\end{table}
For comparison, the parameters for the fit with threshold
$T_\mathrm{th}=3a$ have also been included. The results seem to
be more stable than those from the fits to unsmeared Wilson
loop data (cf.\ Table~\ref{tab:wp1}). This once again signals
the larger ground state overlap of the Coulomb dressed
$Q\bar{Q}$ states as compared to the $q\bar{q}$ states linked
by an unsmeared Wilson line.

Concluding this subsection we reemphasise the striking fact that
Gribov copies seem to have no noticeable effect on the inter-quark
potential. As a result, possible contamination by excited states
notwithstanding, it is fair to say that the potentials calculated
from unsmeared Wilson loops and from Coulomb dressed quark states
do agree.

\section{On the impact of Gribov copies }

In the last section we have shown that the static potentials
obtained from finite length, Coulomb dressed Polyakov lines and
from unsmeared Wilson loops agree. Our calculations have required
a substantial processing of raw numerical data such as the
subtraction of divergent self-energies. The question therefore
arises whether the agreement between standard and Coulomb gauge
calculations also holds for the \emph{bare} finite time
amplitudes, $W(r,T)$ and $C(r,T)$ from (\ref{eq:wp2}) and
(\ref{eq:cp3}), respectively.

If this were true, we would be in a much better position to
understand the impact of Gribov copies as this issue could be
properly addressed before renormalisation.

\subsection{Gribov noise }

Let us start with a description of the actual numerical procedure
which is used to calculate the expectation value of an
observable $O[U]$ in Coulomb gauge. Using standard Monte-Carlo
techniques, a link configuration $\{U\}$ is generated according to
the gauge invariant probability distribution $p[U]$, cf.\
(\ref{eq:i3}). Without gauge fixing an expectation value is
calculated as
\be
  \bra O \ket = \int \mathscr{D}U \, p[U] O[U] \; .
\en
If we evaluate this after an arbitrary gauge transformation, $U
\to {^\Omega}U$, using invariance of the probability measure
$\mathscr{D}U  p[U]$,
\be
  \bra O \ket = \int \mathscr{D}U \, p[U] O[{^\Omega}U] =
  \int \mathscr{D}U \, p[U] \int \mathscr{D}\Omega \, O[{^\Omega}U] \;
  ,
\en
we obviously obtain zero whenever the projection $\int
\mathscr{D}\Omega \, O[{^\Omega}U]$, i.e.\ the group average of
$O$, vanishes. For gauge \textit{variant} field combinations this
is what typically happens.

However, the situation will be different after gauge fixing where,
as a result of the gauge fixing algorithm, any seed configuration
will change into a Coulomb gauge configuration
\be \label{CHANGE}
  \{U\} \; \longrightarrow \; \big\{ {^h} U \big\} \; , \quad h = h[U] \; ,
\en
with (yet unknown) probability $P[ ^h U ]$ characterising the
algorithm. We emphasise that the proper way to interpret the
transformation (\ref{CHANGE}) is as a \textit{change of variables}
\cite{christ:1980b} implying curvilinear coordinates and a metric
formulation which, however, will not be needed here.

It will be sufficient to \textit{define} the expectation value of
an operator $O[U]$ in Coulomb gauge according to
\be
  \langle \langle O \rangle \rangle \; \equiv \; \int
  \mathscr{D} ^h U \; P \big[ \, {^h} U \big] \; O \big[ \, ^h U \big] \; ,
  \label{eq:ii1}
\en
where the double bracket emphasises that $O[U]$ need not be gauge
invariant. Note, however, that the dressed operator $O \big[ {^h}
U \big]$ is, by construction, invariant up to Gribov copies as $h$
always picks an orbit representative. If there were no Gribov
noise the chosen representative would be unique. This uniqueness
is corrupted by the unavoidable randomness inherent in the gauge
fixing algorithm implying a \textit{distribution} of
representative copies. It is this distribution which we want to
study in the following.

To this end let us rewrite the expectation value (\ref{eq:ii1}) in
terms of the Yang-Mills functional integral. The probability of
picking a particular copy $^h U$ is given by the gauge group
average
\be
  \pi \big( {^h} U \big \vert \, U \big) \; \equiv \; \int
  \mathscr{D} \Omega \;
  \rho \big( {^\gt} U \big\vert \, ^\Omega U \big) \; ,
  \label{eq:ii2}
\en
which is gauge invariant by \textit{fiat},
\be \label{PIINV}
  \pi \big( {^h}U \vert \, ^{\Omega}U \big)
  = \pi \big( {^h}U  \big\vert U \big) \; .
\en
The gauge dependent quantity $\rho \big( {^\gt}U \big\vert {^\Omega} U
\big)$ in (\ref{eq:ii2}) denotes the (gauge dependent) probability
distribution for obtaining the output copy ${^\gt}U$ when the
numerical procedure was initiated with input $^\Omega U$ and hence
characterises the gauge fixing algorithm. With these
prerequisites, we find for the distribution $P$ in (\ref{eq:ii1}):
\be
  P \big[ {^h}U \big] \; = \; \int \mathscr{D} U \; \pi \big(
  {^h}U \big\vert U) \; p[U] \; . \label{eq:ii3}
\en
Inserting this into (\ref{eq:ii1}) the expectation value becomes
\be
  \langle \langle O \rangle \rangle \; = \; \int \mathscr{D} U
  \; p[U] \; \int \mathscr{D} {^h}U \; \pi \big( {^h}U \big\vert U
  \big) \;  O\big[ {^h}U \big] \; . \label{eq:ii4}
\en
Before we proceed further let us check that we recover the
familiar formula under the assumption that there is no Gribov
ambiguity. In this case, the gauge fixing algorithm would map $U$
to the unique Coulomb gauge representative ${^{\gt_0}}U $ implying a
sharp distribution
\be \label{PIDELTA}
  \pi \big( {^h}U \big\vert U \big) \; = \; \delta \big( {^h}U
  \big\vert \, ^{\gt_0}U \big) \; , \quad \mbox{where} \quad \gt_0 = \gt_0 [U]
  \; .
\en
In this case, (\ref{eq:ii4}) results in the expression,
\be
  \langle \langle O \rangle \rangle \; = \; \int \mathscr{D}
  U \; p[U] \; O \big[ {^{\gt _0}} U \big] \; , \label{eq:ii5}
\en
which is gauge independent by the (assumed) uniqueness of
$\gt_0[U]$.

\vskip 0.3cm
In the generic case, when Gribov copies are present,
(\ref{PIDELTA}) is no longer true. Nevertheless, gauge invariance
is maintained due to the invariance property (\ref{PIINV}) of $\pi
\big( {^h}U \big\vert U \big)$. This allows for the definition of
a gauge invariant average over Gribov copies,
\be
  \bar{O}[U] \; \equiv \; \int  \mathscr{D} {^h}U \; \pi \big(
  {^h}U \big\vert U \big) \; O \big[ {^h}U \big] \; ,
  \label{eq:ii6}
\en
which indeed satisfies $\bar{O} \big[ {^\Omega}U \big] = \bar{O}[U]$.
Finally, upon plugging this into (\ref{eq:ii4}), the expectation
value may be compactly written as
\be
  \langle \langle O \rangle \rangle \; = \; \langle
  \bar{O} \rangle \; , \label{eq:ii7}
\en
where the brackets on the right-hand side indicate the standard
average with the Yang-Mills probability distribution, $p[U]$.

Let us apply this now to the finite length Polyakov loop
correlator, i.e.\
\be
  O[U] \; \equiv \; C [U] \; = \tr (
  P_T(\vc{x}) \; P^\dagger_T(\vc{y}) ) \; , \label{eq:i12}
\en
and analyse its gauge invariant average $\bar{C}[U]$ associated
with Coulomb gauge.  Since $C [U]$ is gauge dependent it could
nevertheless happen that this average (`Gribov mean'),
\be
  \bar{C} [U] \; \equiv \; \int  \mathscr{D} {^h}U \; \pi \big(
  {^h}U \big\vert U \big) \; C \big[ {^h}U \big]
  \; ,
  \label{eq:i16}
\en
though gauge invariant, actually vanishes when
the copies  are averaged over. We will show shortly that this
is not the case. Therefore, $\bar{C}[U] $ can indeed be
interpreted as a nontrivial gauge invariant observable.

\begin{figure}[h]
\begin{center}
\includegraphics[width=6cm]{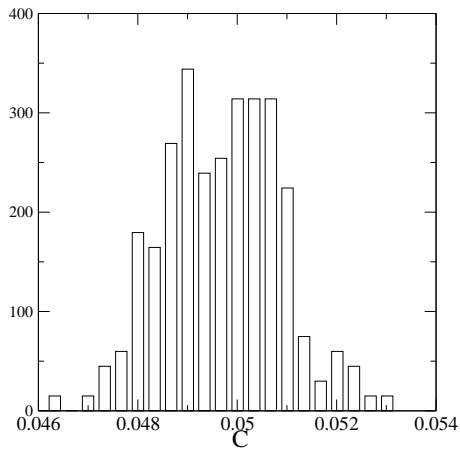}
\caption{ \label{fig:i1} Gribov noise distribution of the
correlator $C$ from (\protect\ref{eq:i12}) for one sample
configuration $U$ and distances $r=3a$ and $r=4a$.}
\end{center}
\end{figure}

Let us consider a specific example. For a $16^4$ lattice and
$\beta =2.4$ we have generated a single sample configuration to
which we apply a random gauge transformation. The result, call it
$U$ is submitted to the gauge fixing algorithm which produces a
Gribov copy configuration, $^{h_1}U$. With this configuration, we
calculate the correlator
\be
  C_1 \; \equiv \; \frac{1}{N^3} \sum _{\ssvc{x}} \tr \;
  ^{h_1}  P_T(\vc{x}) \; ^{h_1} P^\dagger _T(\vc{y}) \, ,
\en
where $\vc{y} = \vc{x} + r \, \vc{e}_3$. We choose the
distances $r=3a$ and $T=4a$ for the present example. Repeating
this procedure $200$ times, leaves us with observables $C_i$,
$i=1 \ldots 200$. Their numerical values are distributed as
shown in FIG.~\ref{fig:i1}.
Note that for a gauge invariant (i.e.\ noise free) field
combination, such as the plaquette, one would find a delta
function distribution according to (\ref{PIDELTA}). However, for
$C[U]$ in (\ref{eq:i16}) we find a broad distribution with mean
$\mu$ and standard deviation $s$ given by:
\be \label{MUS}
  \mu  =  0.0498(1) \ , \quad \mathrm{and} \quad s  =  0.0011(1) \, .
\en
A bootstrap analysis was used for the error estimate in these
values. As they depend significantly on the gauge fixing algorithm
as well as the observable under consideration their interpretation
in terms of physical quantities is somewhat limited. They do,
however, provide information on the impact of Gribov copies since
(i) $\mu$ significantly differs from zero and (ii) the relative
width $s/\mu$ of the distribution is only of the order of 2\%.
This rules out the possibility that the average over Gribov copies
induces a null result.

\subsection{Gauge invariant signals from Gribov-noisy data  }

As a first step, we will investigate the relation between the data
from finite length Polyakov lines and gauge invariant Wilson
loops. Again using the $16^4$ lattice and $\beta=2.4$, we have
generated $10$ seed configurations $U$. For each seed
configuration, we produced $10$ Gribov copies of it. We have then
calculated $r=3a$ and $T=4a$ Wilson loops which are averaged over
the spatial coordinates. The result is called $W[U]$. For each
Gribov copy, the spatial average of finite length Polyakov lines
was also obtained (denoted as $C[^{h}U ]$). As in the previous
subsection, there will be a distribution of values of $C[^{h}U ]$
because of the Gribov noise.

\begin{figure}
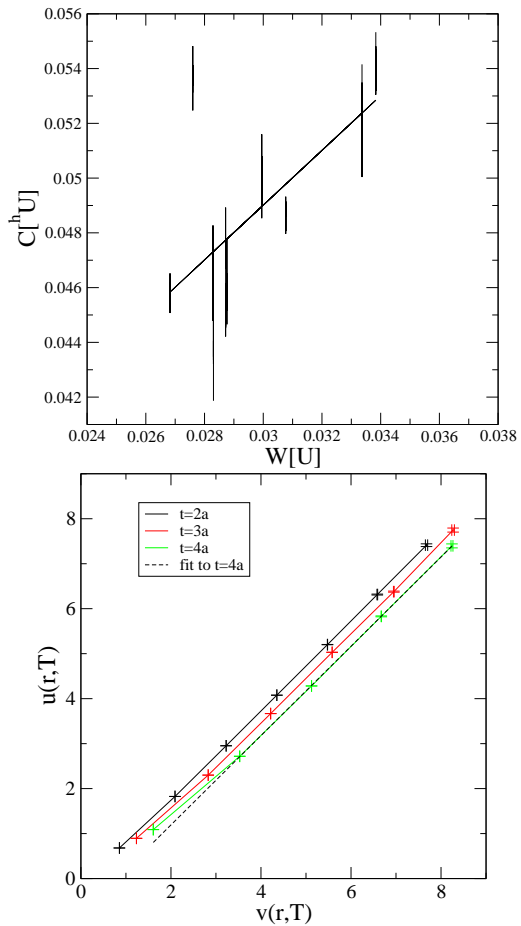

\includegraphics[width=6.8cm]{scatter16}

\includegraphics[width=6cm]{log_w_u}

\caption{ \label{fig:i12}  Scatter plot for ten seed
configurations $U$ (upper panel). Scatter plot of $u(r,T)$, see
(\ref{eq:cp3}), versus  $v(r,T)$, see (\ref{eq:wp2}), for
several values of $r$ and $T$ (lower panel). }
\end{figure}

This data is shown in FIG.~\ref{fig:i12}. For each seed
configuration (and therefore for each value $W[U]$) there is a
band of values for $C[^{h}U ]$ produced by the Gribov copies.
If we assume a perfect correlation of the expectation values,
i.e.
\be
  \langle\langle C \rangle\rangle \; \propto \; \langle W \rangle \, ,
  \label{eq:i17}
\en
the data would be symmetrically scattered around the solid line
also shown in FIG.~\ref{fig:i12}. If, on the other hand, the
Gribov noise were uncorrelated to the underlying seed
configuration, the scatter plot would be homogeneous. Our
findings, shown in FIG.~\ref{fig:i12} left panel, suggest that
a significant correlation survives the Gribov average.

To further test this correlation between finite length Polyakov
lines and gauge invariant Wilson loops, we compare $v(r,T)$
from (\ref{eq:wp2}) as obtained from unsmeared Wilson loops,
with $u(r,T)$ from (\ref{eq:cp3}). FIG.~\ref{fig:i12}, lower
panel, shows the corresponding scatter plot.
We clearly observe a linear correlation between the two
quantities:
\be
  u(r,T) \; = \; v(r,T) \; + \; \epsilon (T) \; . \label{eq:i1111}
\en
The important observation is that $\epsilon (T)$ does not
contribute to the $r$ dependent part of the static potential.
Hence, we find that only the unobservable offset of the potential
is changed if finite length Polyakov lines are employed rather
than unsmeared Wilson lines.
This observation has far reaching consequences: Gribov copies will
affect the quark self-energies, but their impact on the $r$
dependence of the static potential is limited.

\subsection{The Gribov gallery}

In order to trace out the origin of the ideal correlations
discovered in the previous subsection, let us visualise the
spatial dependence of the Gribov copies. This dependence can
change the $r$ dependence of the finite length Polyakov line
correlator as follows. We note that a similar investigation for
the case of Landau gauge was performed
in~\cite{deForcrand:1991ux}. Generally two different runs of
the gauge fixing algorithm result in two different maxima along
the orbit of a seed configuration, $U_\mu$. This in turn yields
two Gribov copies, say $^\gt U_\mu $ and ${^{c \gt}}U_\mu $,
which are related by a residual copy gauge transformation, $c$.
Knowing that the Coulomb gauge (\ref{eq:gf1}) has a trivial
residual gauge freedom consisting of purely time dependent
gauge transformations (\ref{RESGT}), we need to distinguish
between genuine copies $c\gt$ and those of the form $\Omega(t)
h(x)$ which would also satisfy the Coulomb gauge condition. We
therefore define
\be
  \tilde{c}(x)  \; = \; \Omega(t) c(x) \; ,
\en
and use the purely time dependent degree of freedom $\Omega(t)$ to
bring $\tilde{c}$ as close to unity as possible,
\be
  \sum_{\ssvc{x}, t} \tr \tilde{c}(\vc{x}, t) \stackrel{\Omega (t)}{\longrightarrow } \;
  \hbox{max} \, . \label{eq:gf2}
\en
This should suppress the contamination by trivial copies. In other
words, a space dependent $\tilde{c}(x)$ with $ \tr \tilde{c}(x)
/2$ not being $1$ everywhere signals the presence of a non-trivial
Gribov copy. With these prerequisites we can now investigate how
such a copy affects the finite length Polyakov loop correlator.
\begin{figure}
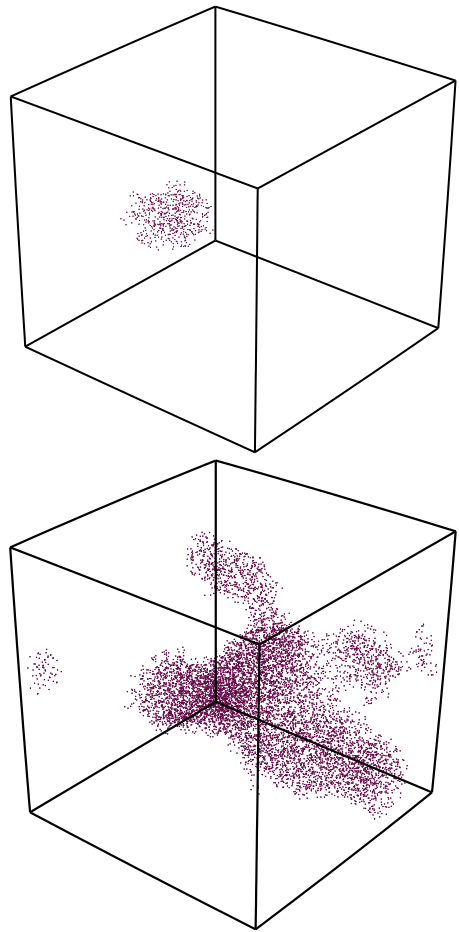

\includegraphics[width=6cm]{out16}

\includegraphics[width=6cm]{out24}

\caption{\label{fig:gribov}Region of space where $\tr \tilde{c}
(x)/2$ resulting from (\ref{eq:gf2}) significantly deviates
from $1$, both on a $16^4$ lattice (upper panel) and on a
$24^4$ lattice (lower panel)}.
\end{figure}

Defining the spatial pure gauge link,
\be
 M(\vc{y}\vc{x}, t) \equiv \tilde{c}^\dagger (\vc{y}, t) \;
 \tilde{c}(\vc{x}, t) \; ,
\en
the Polyakov line correlator associated with the copy
${^{c\gt}}U_\mu$ may be written as
\begin{eqnarray}
  C_{\ssvc{x}\ssvc{y}} [{^{c\gt}}U] &=& \langle \tr \, {^{c\gt}} P_T (\vc{x})
  \; {^{c\gt}} P_T^\dagger (\vc{y}) \rangle \nn \\
  &=&
  \langle  \tr \, M(\vc{y}\vc{x}, 0) \; ^\gt P_T (\vc{x}) \; M^\dagger
  (\vc{y}\vc{x}, T) \;  ^\gt P_T^\dagger(\vc{y}) \rangle \; ,\nn \\
\end{eqnarray}
This means that, whenever the matrices $\tilde{c}(x)$ differ
significantly from unity, an additional non-local correlation is
induced to the correlator of finite length Polyakov lines.

FIG.~\ref{fig:gribov} illustrates the space dependence of
$\tilde{c}(x)$ for two generic lattice configurations for
$\beta =2.4$ at a fixed time slice. The density of points in
the graph is a direct measure for the deviation of $\tr
\tilde{c}/2 $ from $1$. Although the underlying seed
configuration is UV noisy, $\tr \tilde{c} $ changes rather
smoothly throughout space. This explains why $M(\vc{y}\vc{x})$
is not too different from unity for moderate distances $\vert
\vc{y} -\vc{x} \vert $. Note that $M(\vc{y}\vc{x})$ strongly
depends on the efficiency of the gauge fixing algorithm in
singling out a unique maximum on the gauge orbit. Without
adapting the algorithm, there will be also a substantial
dependence of  $M(\vc{y}\vc{x})$ on the volume since the
algorithms generically become inefficient for large volumes.
This fact can be seen from FIG.~\ref{fig:gribov} where two
different volumes have been used in conjunction with the same
gauge fixing algorithm. For the larger volume (lower panel),
the modulation induced by $M(\vc{y}\vc{x})$ is much more
pronounced.

An attractive feature of the Coulomb gauge is the fact that the
Gribov link, $M(\vc{y}\vc{x},t)$,  is purely spatial (i.e.\
located in a single time slice $t$) and thus does not interfere
with time evolution and the transfer matrix formalism. Hence, as
long as the field combination $C[^{h}U ]$ does not vanish upon
averaging over Gribov copies, this average only affects the
definition of the interpolating trial state, $| Q_{\ssvc{x}}
\bar{Q}_{\ssvc{y}} \ket$. In other words, different gauge fixing
algorithms will correspond to different \textit{ans\"atze} for the
quark-antiquark trial states which might differ in their overlap
with the true ground state, $| 0_{\ssvc{x}\ssvc{y}} \ket$. This is
somewhat reminiscent of the situation arising in the process of
smearing Wilson lines.

\section{Conclusions and discussion}

In this paper we have developed a formalism based on Coulomb gauge
fixing which allows us to incorporate the ubiquitous Gribov copies
in a systematic and gauge invariant way. This has been achieved by
defining, for the first time, a Gribov average of gauge variant
field combinations.

We have applied this to a thorough lattice investigation of
Polyakov lines of finite temporal extent introduced by
Marinari et al.~in~\cite{Marinari:1992kh} and further
studied in~\cite{Greensite:2003xf,Greensite:2004ke,Nakamura:2005ux,Nakagawa:2006fk}.
These lines, even if
traced, are not often studied on the lattice since their
expectation value, without gauge fixing, is zero due to their lack
of gauge invariance. Using our formalism, we have shown that, and
explained why, they have a physical interpretation when working in
Coulomb gauge. This has enabled us to extract the confining
inter-quark potential and to analyse the Gribov copies associated
with that gauge and their impact on the overlap with the ground
state in the quark-antiquark sector.

We have seen that the value of the string tension obtained from
the correlator of finite length, dressed Polyakov lines lies
below that obtained from our simulations with unsmeared Wilson
loops (compare Tables 1 and 2). This indicates that the Coulomb
dressed Polyakov lines are, as we expected, less contaminated
by overlaps with higher energy states. Indeed, even using
shorter lines in this approach gives a smaller string tension
than for unsmeared Wilson loops of larger temporal extent.

Many topics brought to light here deserve further exploration. In
particular, Gribov copies have been demonstrated not to hinder the
measurement of the full quark-antiquark potential in this
construction. They merely contribute to the definition of the
interpolating quark-antiquark state, and we have demonstrated in
this paper that its overlap is even better than that of the state
where quark and antiquark are joined by a thin flux line. It would
be interesting to study in a more systematic way the contributions
of excited states to the finite length Polyakov lines.
 There is also an obvious extension of this, namely to use other gauges
and so test whether, as we predict, other dressings have a poorer
overlap with the true ground state of Yang Mills theory.

The Gribov induced non-locality seen in section~4.3 hinders a
proper definition of the constituent quark picture. We would
argue that our results support the, expected, breakdown of a
constituent picture of quarks and antiquarks in the confining
region. The structure of the copies displayed in
FIG.~\ref{fig:gribov}, their volume dependence and sensitivity
to algorithms therefore  merit further investigation. It would
also be very interesting to probe the Gribov non-locality in
the high temperature deconfinement regime where the constituent
quark picture is expected by many to make sense. In this
context, it would also be important to study the relation
between the Gribov non-locality and the confining vortices
advocated in, e.g.~\cite{Greensite:2003bk}. If the confining
vortices are removed, see~\cite{Greensite:2003bk}, do the
Gribov copies seen in FIG.~\ref{fig:gribov} vanish with them
thereby giving rise to a constituent quark picture?

\acknowledgments During the course of this work we benefited
from discussions with Anton Ilderton and Andreas Wipf. We thank
Philippe de Forcrand for correspondence. We are also indebted
to Wolfgang Lutz for a careful reading of the manuscript. The
numerical simulations were carried out using the PlymGrid
facility at the University of Plymouth.



\end{document}